\documentclass [12pt]{article}
\usepackage{graphicx}

\begin{document}
    \begin{titlepage}
    \begin{flushright}
    CARLETON/OPAL PHYS 99-01 \\ February 5, 1999\\
    \end{flushright}
    \vskip 1.5 in
    \begin{center}{\LARGE\bf Confidence Level Computation
         for Combining Searches with Small Statistics }
    \end{center}
    \vskip 0.3 in
    \begin{center} Thomas Junk \\ {\it Carleton University} \\ {\it 1125 Colonel By Drive} \\
    {\it Ottawa, Canada, K1S 5B6}\end{center}
    \vskip 0.3 in
    \begin{abstract}
This article describes an efficient procedure for computing approximate
confidence levels for searches for new particles where the expected
signal and background levels 
are small enough to require the use of Poisson statistics.  The results of many
independent searches for the same particle may be combined easily, regardless
of the discriminating variables which may be measured for the candidate events.
The effects of systematic uncertainty in the signal and background models
are incorporated in the confidence levels.
The procedure described allows efficient computation of expected confidence levels.
\end{abstract}
\thispagestyle{empty}
\end{titlepage}

\section{Introduction}
\label{introsection}

The problem of combining the results of several independent searches
for a new particle and producing a confidence level (CL) has become
very important at the LEP collider in its high-energy phase of running.
Typically, both the expected number of signal events and the expected
number of background events are small, and few candidate events are observed
in the data for any particular search analysis.  The ability to exclude
the presence of a possible signal at a desired CL is often
improved significantly
by combining the results of several searches, particularly if the
sensitivity is limited by the collected luminosity, and not
by a kinematic boundary.  In addition, sophisticated search
analyses may provide information about the observed candidates, such
as one or more reconstructed masses or other experimental information relating
to the expected features of the signal.  
These variables provide better discrimination of
signal from background, and also help to indicate which signal hypothesis is
preferred among many.  Sometimes no such information is available,
and these search analyses must be combined with other types
of analyses for an optimal CL.  Binning the search results of the
analyses in their discriminant variables and treating each bin as
a statistically independent counting search provides a simple,
uniform representation of the data well suited for combination.

Often, as is the case with searches for MSSM Higgs bosons at LEP2,
a broad range of model parameters which affect the production of signal
events must be considered and exclusion limits placed for all possible
values of these parameters.  The expected experimental
signatures of the new particles in general vary with
the model parameters which govern their production and decay, and the
combination of complementary channels provides the best exclusion
for all values of the parameters.  A rapid procedure for computing
confidence levels is therefore necessary in order to explore fully
the possibilities of the model.

This article describes an efficient, approximate method of computing
combined exclusion confidence levels in these cases, allowing also
for the possibility of uncertainty in the estimated signal and background.

\section{Modified Frequentist Confidence Levels}

For the case of $n$ independent counting search analyses, one may
define a test statistic $X$ which discriminates signal-like
outcomes from background-like ones.  An optimal
choice for the test statistic is the 
likelihood ratio~\cite{STATBOOK1,STATBOOK2,ATS2}.
If the estimated signal in the $i^{{\mathrm t\mathrm h}}$
channel is $s_i$, the estimated background is $b_i$,
and the number of observed candidates is $d_i$, then the likelihood
ratio can be written as
\begin{equation}
X = \prod_{i=1}^n X_i,
\end{equation}
with
\begin{equation}
X_i = \left.\frac{e^{-(s_i+b_i)}(s_i+b_i)^{d_i}}{d_i!}\right/
\frac{e^{-b_i}b_i^{d_i}}{d_i!}.
\label{likelihoodratioequation}
\end{equation}
This test statistic has the properties that the joint test statistic
for the outcome of two channels is the product of the test statistics
of the two channels separately, and that it increases monotonically
in each channel with the number of candidates $d_i$.

The confidence level for excluding the possibility of 
simultaneous presence of
new particle production and background (the $s+b$ hypothesis), is
\begin{equation}
CL_{s+b} = P_{s+b}(X\leq X_{obs}),
\label{clsbdefinition}
\end{equation}
{\it i.e.}, the probability, assuming the presence of both signal
and background at their hypothesized levels, that the test statistic
would be less than or equal to that observed in the data.  This
probability is the sum of Poisson probabilities
\begin{equation}
P_{s+b}(X\leq X_{obs}) = \sum_{X(\{d^\prime_i\})\leq X(\{d_i\})}\prod_{i=1}^n
\frac{e^{-(s_i+b_i)}(s_i+b_i)^{d_i^\prime}}{d_i^\prime!},
\label{clsbequation}
\end{equation}
where $X(\{d_i\})$ is the test statistic computed for the observed set of candidates
in each channel $\{d_i\}$, and the sum runs over all possible final outcomes
$\{d_i^\prime\}$ which have test statistics less than or equal to the observed one.
  
The confidence level $(1-CL_{s+b})$ may be used to quote exclusion limits although
it has the disturbing property that if too few candidates are observed
to account for the estimated background, then any signal, and even
the background itself, may be excluded at a high confidence level.
It nonetheless provides exclusion of the signal at exactly the confidence
level computed.  Because the candidates counts are integers, only a discrete
set of confidence levels is possible for a fixed set of $s_i$ and $b_i$.

A typical limit computation, however, involves also computing the
confidence level for the background alone,
\begin{equation}
CL_{b} = P_{b}(X\leq X_{obs}),
\label{clbdefinition}
\end{equation}
where the probability sum assumes the presence only of the background.
This confidence level has been suggested to quantify the confidence of
a potential discovery, as it expresses the probability that background
processes would give fewer than or equal to the number of candidates
observed.
Then the Modified Frequentist confidence level $CL_s$ is computed as the
ratio
\begin{equation}
CL_s = CL_{s+b}/CL_b.
\label{clsdefinition}
\end{equation}
This confidence level is a natural extension of the 
common single-channel CL=1-$CL_s$~\cite{HELENE,PDGBEFORE98},
and for the case of a single counting channel is identical to it.

The task of computing confidence levels for experimental searches with
one or more discriminating variables measured for each event reduces
to the case of combining counting-only searches by binning each search
analyses' results in the measured variables.  Each bin of, {\it e.g.}, the
reconstructed mass, then becomes a separate search channel to be combined
with all others, following the strategy of~\cite{L3MC} and the neutrino-oscillation
example of~\cite{COUSINSFELDMAN}.
In this case, the expected signal in a bin of the reconstructed
mass depends on the hypothesized true mass of the particle and also on the
expected mass resolution.  If the error on the reconstructed mass varies from
event to event such that the true resolution is better for some events and
worse for others, then the variables $s$, $b$, and $d$ may be binned in both
the reconstructed mass and its error to provide the best representation of the available
information.  By exchanging information in bins of the measured variables, different
experimental collaborations may share all of their search result information
in an unambiguous way without the need to treat the measured variables in any
way during the combination.

For convenience, one may add the $s_i$'s, the $b_i$'s, and the $d_i$'s of channels
with similar $s_i/b_i$ and retain the same optimal exclusion limit, just as the
data from the same search channel may be combined additively for running periods
with the same conditions.    The same search with a new beam energy or other
experimental difference should of course be given its own set of bins (which
may be combined with others of the same $s_i/b_i$).

\section{Confidence Level Calculation}
\label{calcsection}

  The task of summing the terms of Equation~\ref{clsbequation} can be formidable.
For $n$ channels, each with $m$ possible outcomes, there are ${\cal O}(n^m)$ terms
to compute.  This sum is often carried out with a 
Monte Carlo~\cite{L3MC,COUSINSFELDMAN}, selecting representative
outcomes of the experiment and comparing their test statistics with the test statistic
computed with the data candidate event counts.
Another alternative, described in this article, is to compute
 the probability distribution function
(PDF) for the test statistic for a set of channels, and iteratively combine additional
channels by convoluting with the PDFs of their test statistics.

  The PDF of the test statistic for
a single channel is a sum of delta functions at the accessible values of $X_i$.
These may be represented as a list of possible outcomes
\begin{equation}
{(X_i^j,p_i^j)},
\label{plistequation}
\end{equation}
where $X_i^j$ is the test statistic for the $i^{{\mathrm t\mathrm h}}$ channel if it were to
have $j$ events, and $p_i^j$ is the Poisson probability of selecting $j$ events
in the $i^{{\mathrm t\mathrm h}}$ channel if the underlying average expected rate is $s_i+b_i$
when computing $CL_{s+b}$, or only $b_i$ when computing $CL_b$.  The list
is formally infinitely long, but
one may truncate it when the total probability sum of the outcomes in the
list exceeds a fixed quantity, or one may select all $j$ such that $X_i^j\leq X_{obs}$.

For the case of two channels, one forms the probabilities and test statistics for
the joint outcomes multiplicatively,
\begin{equation}
{(X_i^jX_{i^\prime}^{j^\prime},p_i^jp_{i^\prime}^{j^\prime})},
\end{equation}
to form a representation of the PDF of the test statistic for the joint outcomes
of two channels.  One may then iteratively combine all channels together and use
the list to compute the confidence level by adding the probabilities of outcomes
with test statistics less than or equal to that observed.  This reintroduces the
computational difficulty of enumerating all possible experimental outcomes, and
hence one needs to introduce an approximation to limit the complexity of the problem.

The approximation is to bin the PDF of the test statistic at each combination
step.  The cumulative PDF may be obtained from the listing of outcomes by sorting
them by their test statistics and accumulating the probabilities.  Then fine bins
of the cumulative PDF may be filled with possible outcomes.  A useful binning
covers very small probabilities logarithmically in order to represent
small CL's more exactly, and has a uniform binning for larger probabilities.
The finer the bins, the more precise the computed CL will be; in the limit of
infinitely fine bins, the problem reduces once again to adding the probabilities
of all possible outcomes.

To guarantee a conservative CL for setting limits, one may, at each combination step, record as
a possible experimental ``outcome'' the smallest test statistic within a bin
coupled with the largest accumulated probability within the same bin.
The list now consists of test statistics and the cumulative probability of observing
that test statistic or less, and the differential PDF of $X$ may be recovered
from it.

The process is then repeated iteratively for all channels to be combined.
The running time on a computer is proportional to the number of channels,
the number of bins kept in the PDF of $X$, and increases with the expected
number of events in the channels.  To improve the accuracy of the approximation,
the search channels should be sorted in order of $s_i/b_i$, with the channels
with the largest $s_i/b_i$ combined last.

Once all channels have been combined, the test statistic is computed for the
candidate events observed in the experiment and $CL_{s+b}$, $CL_b$ and $CL_s$
may be computed using Equations~\ref{clsbdefinition}, \ref{clbdefinition} 
and~\ref{clsdefinition}.  Furthermore, the PDFs of $X$ in the signal+background
and background hypotheses allow computation of the expected confidence levels
$\langle CL_{s+b}\rangle$, $\langle CL_b\rangle$, and
$\langle CL_{s}\rangle$, assuming the presence only of background.  These
are indications of how well an experiment would do on average in excluding
a signal if the signal truly is not present, and are the important figures
of merit when optimizing an analysis for exclusion.

When computing $\langle CL_b\rangle$, the outcomes
are already ordered by their test-statistic and only the probabilities
are needed:
\begin{equation}
\langle CL_b\rangle = \sum\limits_{i=1}^{N_{blist}} \left[ p_i^b
\sum\limits_{j=1}^ip_j^b\right],
\end{equation}
where $N_{blist}$ is the number of entries in the table of the PDF of
$X$ for the background-only hypothesis, and $p_j^b$ is the $j^{\mathrm{th}}$
probability in the list, where the test statistic $X$ increases
with increasing $j$.  For total expected backgrounds of more than
about 3.0 events in channels 
with non-negligible sensitivity to the signal,
$\langle CL_b\rangle\approx 0.5$.

The values of $\langle CL_{s+b}\rangle$ and $\langle CL_{s}\rangle$
can be computed similarly, although the PDF of $X$ is needed in the
$s+b$ hypothesis as well as the background-only hypothesis.
\begin{equation}
\langle CL_{s+b}\rangle = \sum_{i=1}^{N_{blist}}\left[ p_i^b
\sum\limits_{X_j^{s+b}\leq X_i^b}p_j^{s+b}\right],
\end{equation}
and
\begin{equation}
\langle CL_{s}\rangle = \sum_{i=1}^{N_{blist}} \left[ p_i^b
\frac{\sum\limits_{X_j^{s+b}\leq X_i^b}p_j^{s+b}}
{\sum\limits_{j=1}^i p_i^b}\right],
\end{equation}
where $p_j^{s+b}$ is the $j^{\mathrm{th}}$ entry in the PDF table
of $X$ for the $s+b$ hypothesis, and $X_j^{s+b}$ is its corresponding
value of $X$.

The difference between this method and that
described by Cousins and Feldman~\cite{COUSINSFELDMAN}
is the choice of test statistic (referred to as
the ``ordering principle'' in~\cite{COUSINSFELDMAN}).  The likelihood ratio
of Equation~\ref{likelihoodratioequation} has the advantages that it
is the most powerful test statistic for distinguishing the $s+b$ hypothesis
from the background-only hypothesis, and also because it does not depend
on the range of possible models of new physics considered when testing a
particular signal hypothesis.  With the test statistic 
of~\cite{COUSINSFELDMAN,WILKS},
a signal hypothesis can be excluded because other signal hypotheses fit the data
better.  The use of the test statistic of~\cite{COUSINSFELDMAN,WILKS}
does not allow the exclusion of the entire model space under study -- one must
be careful to include the null hypothesis of no new particle production in the
space of models to be tested.  In addition, there may be more than one
new physics signal present in the data.  The method of~\cite{COUSINSFELDMAN}
is ideal for the case in which the possible model space is fully known, and
it is known that exactly one of the points in model space corresponds to
the truth.

For purposes of discovery, $1-CL_b$ indicates the probability
that the background could have fluctuated to produce a distribution of candidates
at least as signal-like as those observed in the data.
This probability depends on the signal hypothesis because channels
with small $s_i/b_i$ do not contribute as much to the computation
of $CL_b$ as those with large $s_i/b_i$.  In the case that a particle of
unknown mass is sought, analyses which reconstruct the mass provide
discrimination among competing signal hypotheses when a clear signal
is present, rather than the presence of an excess of candidates.
Nonetheless, the probability in the upper tail of the $X$ distribution
in the $s+b$ hypothesis may be used to exclude a signal hypothesis
because it does not predict enough signal to explain the candidates
in the data.

\section{Systematic Uncertainty on Signal and Background}

\label{syserrors}

The effect on the confidence levels from systematic uncertainties
in the signal estimations $\{ s_i\}$ and background estimations
$\{ b_i\}$ can be accommodated by a generalization of the method
of Cousins and Highland~\cite{COUSINSHIGHLAND}.  This approach
was originally created for one-channel searches with systematic
uncertainty on the signal estimation only.  A very similar approach
for handling background uncertainty is described by C. Giunti 
in~\cite{GIUNTI}.  The generalization of this technique to the
case of many channels with errors on both signal and background
is summarized here.

When forming the list of the probabilities and test statistics of
possible outcomes for a channel,
each entry in the list is affected by the systematic uncertainties
on the signal and background estimations for that channel.  This
effect is computed by averaging over possible values of the signal
and background given by their systematic uncertainty probability 
distributions.  For purposes of implementation, these probability
distributions are assumed to be Gaussian, with the lower tail cut
off at zero, so that negative $s$ or $b$ are not allowed.

When computing the PDF of $X$ for the $s+b$ case, the probability to
observe $j$ events in channel $i$ with estimated signal $s_i\pm\sigma_{s_i}$
and estimated background $b_i\pm\sigma_{b_i}$, is
\begin{equation}
p_i^j = \frac{\displaystyle
         \int_0^\infty{\mathrm d} s^\prime
         \int_0^\infty{\mathrm d} b^\prime
         \frac{e^{-{\left((s^\prime - s_i)^2/2\sigma_{s_i}^2+(b^\prime - b_i)^2/2\sigma_{b_i}^2\right)}}}
              {2\pi\sigma_{s_i}\sigma_{b_i}}
         \frac{e^{-(s^\prime+b^\prime)}(s^\prime+b^\prime)^{j}}{j!}
         }
         {\displaystyle
         \int_0^\infty{\mathrm d} s^\prime
         \int_0^\infty{\mathrm d} b^\prime
         \frac{e^{-{\left((s^\prime - s_i)^2/2\sigma_{s_i}^2+(b^\prime - b_i)^2/2\sigma_{b_i}^2\right)}}}
              {2\pi\sigma_{s_i}\sigma_{b_i}}
         },
\label{systint1}
\end{equation}
which is used in each entry in the list of Equation~\ref{plistequation}.  While the denominator
is a product of error functions, the numerator may be computed numerically.
When computing the PDF of $X$ for the background-only case, the averages are only done over the
background variation.

To extend this to the multichannel case, additionally the test statistic must be averaged over
the systematic variations because it, too, depends on $s_i$ and $b_i$:
\begin{equation}
X_i^j \rightarrow \frac{\displaystyle
         \int_0^\infty{\mathrm d} s^\prime
         \int_0^\infty{\mathrm d} b^\prime
         \frac{e^{-{\left((s^\prime - s_i)^2/2\sigma_{s_i}^2+(b^\prime - b_i)^2/2\sigma_{b_i}^2\right)}}}
              {2\pi\sigma_{s_i}\sigma_{b_i}}
         X_i^j
         }
         {\displaystyle
         \int_0^\infty{\mathrm d} s^\prime
         \int_0^\infty{\mathrm d} b^\prime
         \frac{e^{-{\left((s^\prime - s_i)^2/2\sigma_{s_i}^2+(b^\prime - b_i)^2/2\sigma_{b_i}^2\right)}}}
              {2\pi\sigma_{s_i}\sigma_{b_i}}
         }.
\label{systint2}
\end{equation}
This average is also computed numerically.  It is computed both when the sum over all possible
experimental outcomes is performed and when the test statistic is computed for the data candidates,
ensuring that the data outcome is identical with one of the possible outcomes in the PDF tables.
This is important for confidence levels computed with a single channel, when all outcomes
are listed in the PDF table.

\section{Numerical Examples}

The above algorithm has been tested in a variety of ways.
For general use, a program implementing it is available at \\
{\tt http://home.cern.ch/$\sim$thomasj/searchlimits/ecl.html}.

\begin{figure}
\begin{center}
\includegraphics[width=12cm]{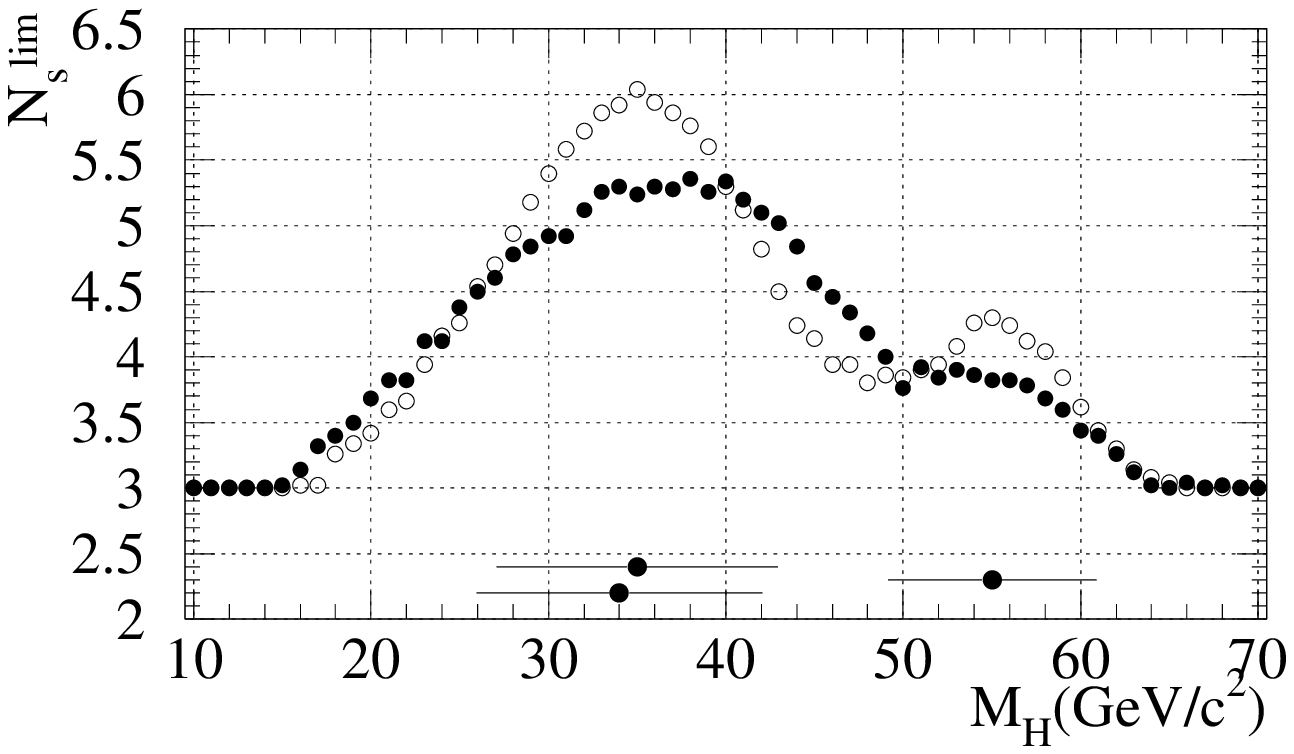}
\caption{The 95\% CL upper bound on the number of events as a function of
a hypothetical Higgs mass, using two test statistics, the likelihood
ratio (filled circles) and events weighted by $s_i/b_i$ (empty circles).
Candidates are shown with their respective mass resolutions at the bottom
of the figure.  The total background is four events expected to be uniformly
distributed from zero to 100 GeV/$c^2$.}
\label{fictive}
\end{center}
\end{figure}

\begin{figure}
\begin{center}
\includegraphics[width=13cm]{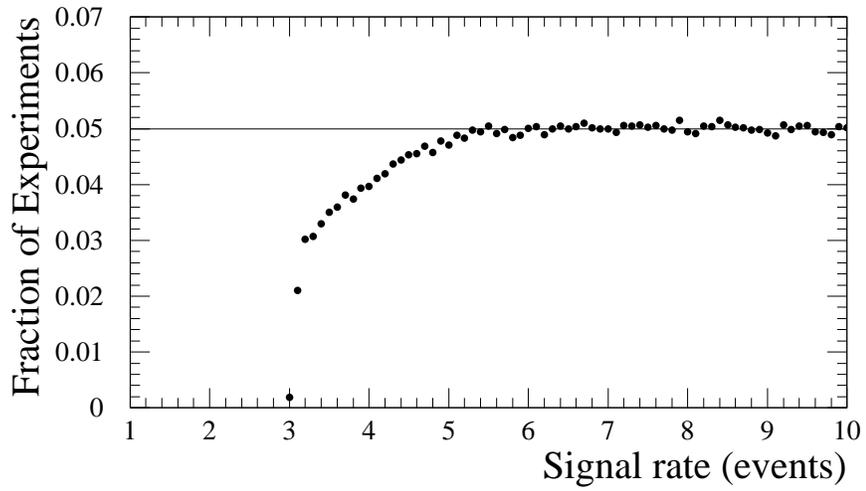}
\caption{The false exclusion rate for the mock Higgs search experiment in the
presence of a real signal at $m_{\mathrm H}$=77 GeV/$c^2$, for 95\% CL computation.
The error bars are hidden within the plot symbols.  If a pure frequentist
approach were taken (using $CL_{s+b}$), then the false exclusion probability would be
flat at 5\%.}
\label{falsexclude}
\end{center}
\end{figure}

\begin{itemize}
\item If a single channel has 3.0 expected signal events, no
expected background events, and no observed candidates,
then $CL_s=4.9787\%$ as expected from an exact
computation.  $CL_b=1.0$ in this case.  For experiments with
few possible outcomes, this technique yields exact CL's.

\item If this single channel is broken up into arbitrarily many
pieces (say, a few hundred), equally dividing up the 3 expected
signal events, each with no background or candidates, the limit
is the same as that for the single channel.

\item  If a channel with no expected signal, but some expected background
(and corresponding data candidates) is added to the combination, then
$CL_s$ is not changed significantly, while $CL_{s+b}$ and $CL_b$ reflect
the relationship between the expected background and the observed candidate
count.

\item A more realistic example requiring the binning of search results
and combination of those bins has been explored by simulating a typical
search for the Higgs boson (or any new particle)
in high-energy particle collisions, where the mass of each
observed candidate may be reconstructed from measured quantities.
The mock experiment has an expected background of 4 events, uniformly
distributed from 0 to 100~GeV/$c^2$ in the reconstructed mass.
The resolution of the reconstructed mass of signal events, were a signal to exist, 
decreases linearly from 10.5~GeV/$c^2$ at $m_H$=10~Gev/$c^2$ to 
3.3~GeV/$c^2$ at $m_H$=80~GeV/$c^2$, where $m_H$ is the mass of
the Higgs boson (or other new particle).  In a real search, the signal resolutions
and background levels are typically obtained from Monte Carlo simulations.
Three candidates were introduced with measured masses of 34, 35, and 55~GeV/$c^2$.  

To explore the limits one may set on Higgs production,
the space of possible values of $m_H$ was explored from 10~GeV/$c^2$ to
70~GeV/$c^2$, and the total expected signal count was studied between
2 and 6.5 events.  For each pair of $m_H$ and the signal count, histograms
of the expected signal and background were formed in fine
bins from 0 to 100~GeV/$c^2$.  The candidates were also histogrammed using the
same binning as the signal and background.  Each bin of these histograms was considered
a separate search channel, and the confidence level $CL_s$ was formed.

The 95\% CL upper limits ($CL_s<0.05$) on the signal $s=\sum_{i=1}^ns_i$
are shown in Figure~\ref{fictive} for two choices of the test statistic $X_i$:
the likelihood ratio of Equation~\ref{likelihoodratioequation}, and the test statistic 
$X_i = d_is_i/b_i$.  This latter test statistic is the event count weighted
by the signal/background ratio, and it is combined additively from channel to channel.

The two test statistics perform differently under these circumstances, and the
method described in this article can be used to evaluate the effects of
changing the test statistic.  The expected confidence levels $\langle CL_{s+b}\rangle$
and $\langle CL_{s}\rangle$ provide discrimination of which test statistic is the
best choice.

\item  The probability coverage of the techinique was explored by testing to
see how often a true signal would be excluded at the 95\% CL.
The same mock experiment as described above was used, but the candidates were
distributed according to a signal+background expectation with signal levels varying
from 3 events to 10 events, with a true mass of 77~GeV/$c^2$.  Many experiments
were simulated with different populations of candidates according to the
hypothesis, and the probability of excluding a true signal, hypothesized
to have the same strength as was used to simulate the experiments,
at 95\% CL is shown in Figure~\ref{falsexclude}.  The exclusion fraction is
smaller than 5\% for low expected signal rates, a consequence of the use of 
the Bayesian
$CL_s=CL_{s+b}/CL_b$, where some of the exclusion power is lost by
dividing by $CL_b$.  Alternatively, one may use $CL_{s+b}$ exclusively,
which would give the proper limit.
In the latter case, the sensitivity $\langle CL_{s+b}\rangle$ should be quoted with
experimental results as well to cover the case of much fewer candidate events
than the background expectation, giving a more stringent limit than would be
warrented by the sensitivity of the experiment.

\item For combining the search results from four LEP experiments for the
MSSM Higgs, nearly 100 separate search analyses from different energies, performed
by different collaborations, have been combined using this technique.  For
a model point with $m_{\mathrm h}$ and $m_{\mathrm A}$ near the exclusion limit for the combined
data from 1997 and before, this method computes $CL_s=5.380\%$, while
an exact computation yields $CL_s=5.332\%$, both corresponding to an exclusion
not quite at the 95\% level.  For this test, the bin width for the 
PDF of $X$ was 0.03\% above probabilities of 1\%, and 20 bins per decade below 1\%.

\item To test the correctness of the strategy for handling systematic uncertainty
in the signal, the results of Table~1 in Reference~\cite{COUSINSHIGHLAND} have
been reproduced.  In all cases, the Monte Carlo confidence levels of 
Reference~\cite{COUSINSHIGHLAND} were reproduced
at least as well as by Equation~(17a) in the same paper.  This equation is
\begin{equation}
 U_n = U_{n0}\left[ 1 + \left\{ 1 - 
\left( 1 - \sigma_r^2E_n^2\right)^{1/2}\right\}/E_n\right],
\label{COUSINSHIGHLAND17A}
\end{equation}
where $U_n$ is the upper limit, including the effects of systematic uncertainty,
on the signal at a desired CL if $n$ candidate events were observed in the
data, $U_{n0}$ is the upper limit on the signal at the
same CL without
the effects of systematic uncertainty, $\sigma_r$ is the relative uncertainty on
the signal ({\it e.g.}, from uncertainty on the efficiency or luminosity),
and $E_n \equiv U_{n0} - n$.  The results of this test are
shown in Table~\ref{COUSINSHIGHLANDREPTABLE}.

\end{itemize}

\section{Limitations}

Because the binning of the PDF of the test statistic $X$ has a finite
resolution, experimental outcomes with very small probabilities of occurring are
not represented correctly.  When using the conservative choice of
filling the bins described above,
these outcomes are overrepresented in the final outcome.  For the purposes
of discovery, however, this approach is not conservative.
When computing the CL for
a potential discovery, one must compute the sum of probabilities of
fluctuations of the background
giving results that look at least as much like the signal as the
observed candidates, or more.
Conversely, one may add up all the probabilities for outcomes less signal-like
than observed and subtract it from unity.  This involves precise accounting
of many outcomes with small probabilities, and the approximation presented
here will not suffice.  The most useful case for this technique is in forming
CL limits near the traditional 90\%, 95\%, and 99\% levels.

Another limitation is that correlations between the systematic uncertainties
of different search channels are not incorporated.  If the results of a search
are binned in a discriminant variable, the signal estimations in neighboring
bins may share common uncertainties, as may the background estimations.  Similarly,
if several experimental collaborations perform similar searches using similar
models for the signal and background, then their results will share common
systematic uncertainties.  A Monte Carlo computation of
the confidence levels is needed when the effects of correlated errors are
expected to be large.  The effect can be estimated by replacing blocks of
correlated parameters $s_i$
and $b_i$ with biased values and recomputing the confidence levels.

The technique described in this article also requires that the value of the
test statistic is defined for each single-bin counting search channel, and that
these test statistics may be combined to form a joint test 
statistic\footnote{The combination rule for the test statistic
needs to be associative in order for the iterative combination of one
search channel to a list of combined results of other search channels
to be well defined.  The combination rule also needs to be commutative
so that the order in which the combination is performed does not affect 
the outcome.}.
More complicated test statistics which cannot be separated into contributions
from independent channels cannot be used with this technique.  A Monte Carlo
approach is suggested in order to use such test statistics.  The likelihood
ratio test statistic of Equation~\ref{likelihoodratioequation}, because it
combines multiplicatively, is well suited for this technique.
 
Special care has to be taken in the case that candidate events can have
more than one interpretation.  A single event may appear in more than one
bin of an analysis or may appear in two separate analyses due to ambiguities
in reconstruction or interpretation.  The most rigorous treatment of
such cases is to construct search bins which
contain mutually exclusive subsets of the search results.  For example, 
one may wish combine three counting channels, A, B, and C, and candidate
events may be classified as passing the requirements of A, B, or C separately,
while some may pass the requirements of both A and B, or both A and C, etc.  In this
case, one would construct seven exclusive classification bins, A, B, C, AB, AC, BC,
and ABC, and proceed as before.
  In general, if a combination
has a total of $n$ bins, then there are $2^n-1$ possible classifications of each event
if multiple interpretations are allowed.  The nature of the analyses will necessarily
reduce the size of this possible overlap problem, and only cases in which significant
overlap is expected for signal or background events need to be considered.

\section{Summary}

An efficient technique for computing confidence levels for exclusion of small signals
when combining a large number of counting experiments has been presented.  The results
of sophisticated channels with reconstructed discriminating variables are binned and the
separate bins are treated as independent search channels for combination.  A variety of
test statistics may be used to evaluate their effects on the confidence levels.  The approximate
confidence levels obtained are very close to the values of computationally intensive direct
summations of probabilities of all final outcomes, or to those obtained by
Monte Carlo simulations, and the
accuracy of the approximation is adjustable.  The confidence levels are either exact or more
conservative than the true values from explicit summation.  Average expected confidence
levels may easily be calculated from the results, and the probability distributions of
the test statistic may be used to construct confidence belts using the techniques described in
Reference~\cite{COUSINSFELDMAN}.  Uncorrelated systematic
uncertainties in the signal and background models are incorporated in a natural manner.
Monte Carlo alternatives are
suggested when the effects of correlated
systematic uncertainties are expected to be large and in the
case of potential discoveries.
This technique is useful for efficiently scanning many possible models for production of signals with
different signatures and combining the results of searches sensitive to these
different signatures.

\clearpage
\newpage

\clearpage
\newpage

\clearpage
\newpage

\begin{table}
\begin{center}
\caption{Reproduction of Table~1 of Reference~\cite{COUSINSHIGHLAND}, together
with the computation of the same quantity using the method of this article.
Listed are the 90\% CL upper limits on the signal for a single counting
measurement with
no background, no uncertainty on the background, and $n$ candidates.  The
relative uncertainty on the signal is $\sigma_r = \sigma_s/s$.  The Monte
Carlo column (MC) is also from Reference~\cite{COUSINSHIGHLAND}.  The missing
entry in the column for Equation~(17a) has a square root of a negative argument,
indicating that the expansion used to derive the formula has reached its
limit of validity.}
\label{COUSINSHIGHLANDREPTABLE}
\begin{tabular}{c c c c c}
  &      &      &      &      \\
\hline
$n$ & $\sigma_r$ & MC & Eq. (17a) & This Work \\
\hline
0 & 0.00 & 2.30 & 2.30 & 2.30 \\
  & 0.10 & 2.33 & 2.33 & 2.33 \\
  & 0.20 & 2.42 & 2.41 & 2.42 \\
  & 0.30 & 2.60 & 2.58 & 2.61 \\
  &      &      &      &      \\
1 & 0.00 & 3.89 & 3.89 & 3.89 \\
  & 0.10 & 3.94 & 3.95 & 3.95 \\
  & 0.20 & 4.13 & 4.14 & 4.14 \\
  & 0.30 & 4.51 & 4.57 & 4.53 \\
  &      &      &      &      \\
2 & 0.00 & 5.32 & 5.32 & 5.32 \\
  & 0.10 & 5.41 & 5.41 & 5.42 \\
  & 0.20 & 5.71 & 5.72 & 5.71 \\
  & 0.30 & 6.30 & 6.78 & 6.32 \\
  &      &      &      &      \\
3 & 0.00 & 6.68 & 6.68 & 6.68 \\
  & 0.10 & 6.80 & 6.81 & 6.81 \\
  & 0.20 & 7.21 & 7.27 & 7.22 \\
  & 0.30 & 8.05 & ---  & 8.05 \\ \hline
\end{tabular}  
\end{center}
\end{table}

\end{document}